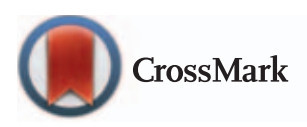

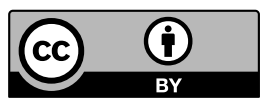

PAPER

# Symmetry/asymmetry within a cylindrical lattice model of nuclear structure predicts cosmic abundance/scarcity

Ray Walsh

Chemistry Dept., College of Western Idaho, Nampa, ID, United States of America

**E-mail:** raywalsh@cwi.edu

## Abstract

Protons and neutrons are the smallest forms of matter with measurable size, and each nucleon has a mix of three up (+2/3) and down (−1/3) quarks. Quarks are treated as point-like particles within polygon-based geometries in order to model the structures of stable nuclides through $^{36}$Ar. The model derives *ab initio* from the proton's radius (r = 0.8414 fm), the hadron's prolate spheroid shape (from the transition to the nucleon's first excited state (the Δ(1232) resonance), and the separation distance between bound nucleons (≈0.8 fm, from the Argonne v18 *NN* potential). The spatial extent of the prolate spheroid nucleon is assumed to arise from its three quarks, which implies a qualitatively linear quark sequence. Spin–spin forces repel the two like-flavored quarks to opposite ends of the prolate nucleon leaving the unlike quark in the middle. It then follows that the quark-to-quark distance within the nucleon corresponds to the nucleon's radius. Nucleons link by quark-to-quark interactions to form proton-neutron short-range correlated pairs (*pn* SRCs), separated by a distance assumed equivalent to the proton's radius. Alternating nucleons thus produce regularly alternating up/down point-like quark sequences. We contemplate various structures for each stable nuclide through $^{36}_{18}$Ar and include the one whose calculated rotational radius (derived from the radius formula of a regular polygon) best correlates with its IAEA-accepted charge radius (r(31) = .98, p<.001). Nucleon alternation inherently makes *pn* SRC pairs, and produces the observed equal numbers of protons and neutrons (Z = N) found within isotopes $^{4}_{2}$He, $^{12}_{6}$C, $^{14}_{7}$N, $^{16}_{8}$O, $^{20}_{10}$Ne, $^{24}_{12}$Mg, $^{28}_{14}$Si, and $^{32}_{16}$S, which together comprise 99.5% of nonhydrogen baryonic matter. Bilateral structural symmetry emerges as a sensitive and specific predictor of cosmic abundance. Opposing deuteron-deuteron' alternating quark charge sequences produce alternating and unequal electromagnetic fields shown capable of modelling the close-range attraction and far-range repulsion of the fusion potential curve and Coulomb potential energy barrier.

## 1. Introduction

Representational models of the atom have a long history of conveying simple but salient structural information while laying the groundwork for more detailed analyses. Examples include Lewis structures and Van't Hoff's tetrahedron model of carbon [1, 2]. Though simplistic, these two representational models were specifically cited by Pauling in the first quantum mechanical treatment of the chemical bond [3].

Here we examine the empirical knowledge of light nuclide structure and composition towards the goal of describing a self-consistent geometric model of nuclides. The factors examined include nucleon size and shape, the sequence and occupancy of nucleons, and cosmic relative abundance. Light nuclide structures that best fit empirical data result when protons and neutrons alternate within a cylindrical lattice comprising stacked 6-nucleon rings. The model's representational coordinate-space geometry is self-consistent, surprisingly straightforward, and comprehensible on its own terms.

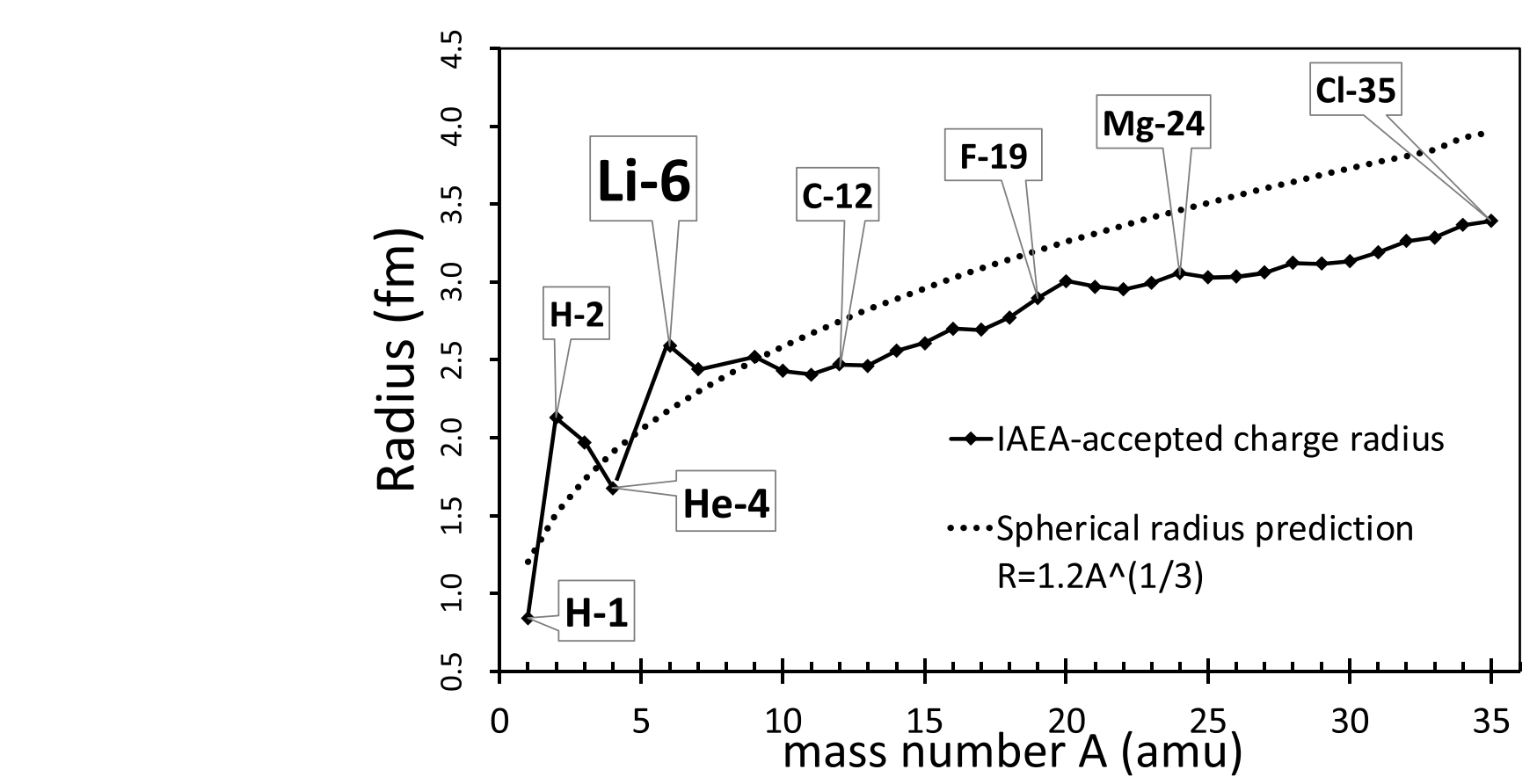

**Figure 1.** The charge radius-to-mass number plot of light nuclides (solid line) is initially erratic but becomes more linear above $^{6}$Li. The assumption of sphericity (dotted line) scales poorly with light nuclide radii. Curiously, the radius of $^{4}$He (1.68 fm) is smaller than $^{2}$H (2.23 fm) but has twice the mass, and the radius of $^{6}$Li (2.59 fm) is oddly larger than $^{12}$C (2.43 fm), which has twice the mass [6].

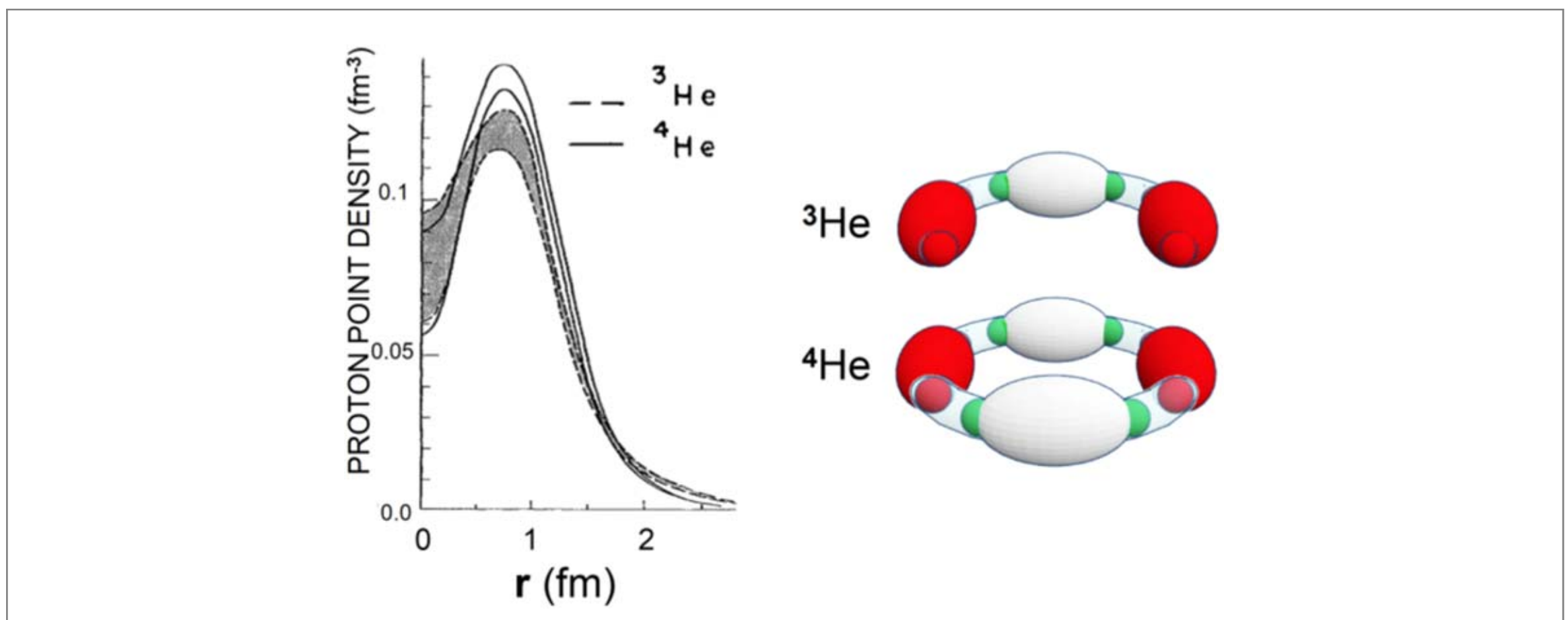

**Figure 2.** The radial plot of the experimental electron-scattering point-like proton densities of $^{3}$He and $^{4}$He (panel left) indicate a lower likelihood of finding a proton near the center (r = 0) of each helium nuclide [9]. The proposed alternating nucleon structures of $^{3}$He and $^{4}$He (panel right) are both empty in the center, consistent with this experimental finding.

Nuclear size, shape, and density are fundamental intrinsic physical properties. The atomic nucleus and its component nucleons are commonly depicted as hard spheres, closely packed within a roughly spherical nucleus. Though perhaps useful pedagogically, empirical evidence suggests that neither nucleons nor nuclei are generally spherical. The electromagnetic excitation of the nucleon to its first excited state, the Δ resonance, has provided clear evidence of a prolate spheroid deformation of the ground state charge distribution of the nucleon, and an oblate spheroid for the excited state [4]. While the charge radii of larger nuclides may scale (roughly) with the cube root of the mass number, only in some cases is this consistent with a spherical nucleus [5]. In fact, the erratic sawtooth radius-to-mass plot of the lightest nuclides below $^{7}$Li is *not* compatible with sphericity (figure 1). Atomic spectroscopy of the deuteron ($^{2}$H) indicates a prolate shape, for example [7, 8]. Furthermore, the close-packing of nucleons does not account for the deuteron radius of 2.13 fm relative to the larger $^{4}$He nucleus of 1.68 fm even though the latter has twice as many nucleons. The $^{12}$C radius (2.43 fm) is oddly smaller than the radius of $^{6}$Li (2.59 fm) although $^{12}$C has twice the nucleons of $^{6}$Li [6]. Strangely, the radial charge and point-proton densities of the stable helium isotopes $^{3}$He and $^{4}$He (figure 2, panel left) suggest a hole or central depression in the nuclide structure [9]. $^{12}$C and $^{16}$O demonstrate a similar central dip in their radial charge densities [10, 11]. The anomalous central dip in charge density of some light nuclides is not consistent with a crystalline lattice of closely packed nucleons, in contrast to the generally constant or flat central charge densities of the medium to heavy nuclides above $^{40}$Ca. Given that 99.9% of the ordinary (baryonic) mass of the universe is contained within the lightest nuclides through $^{36}$Ar (figure 3), these curiosities and inconsistencies are no trivial matter.

In addition to intrinsic structural factors, nucleon knock-out experiments (in which a projectile impacts a target and the momenta of the fragments are compared) have demonstrated an important compositional factor.

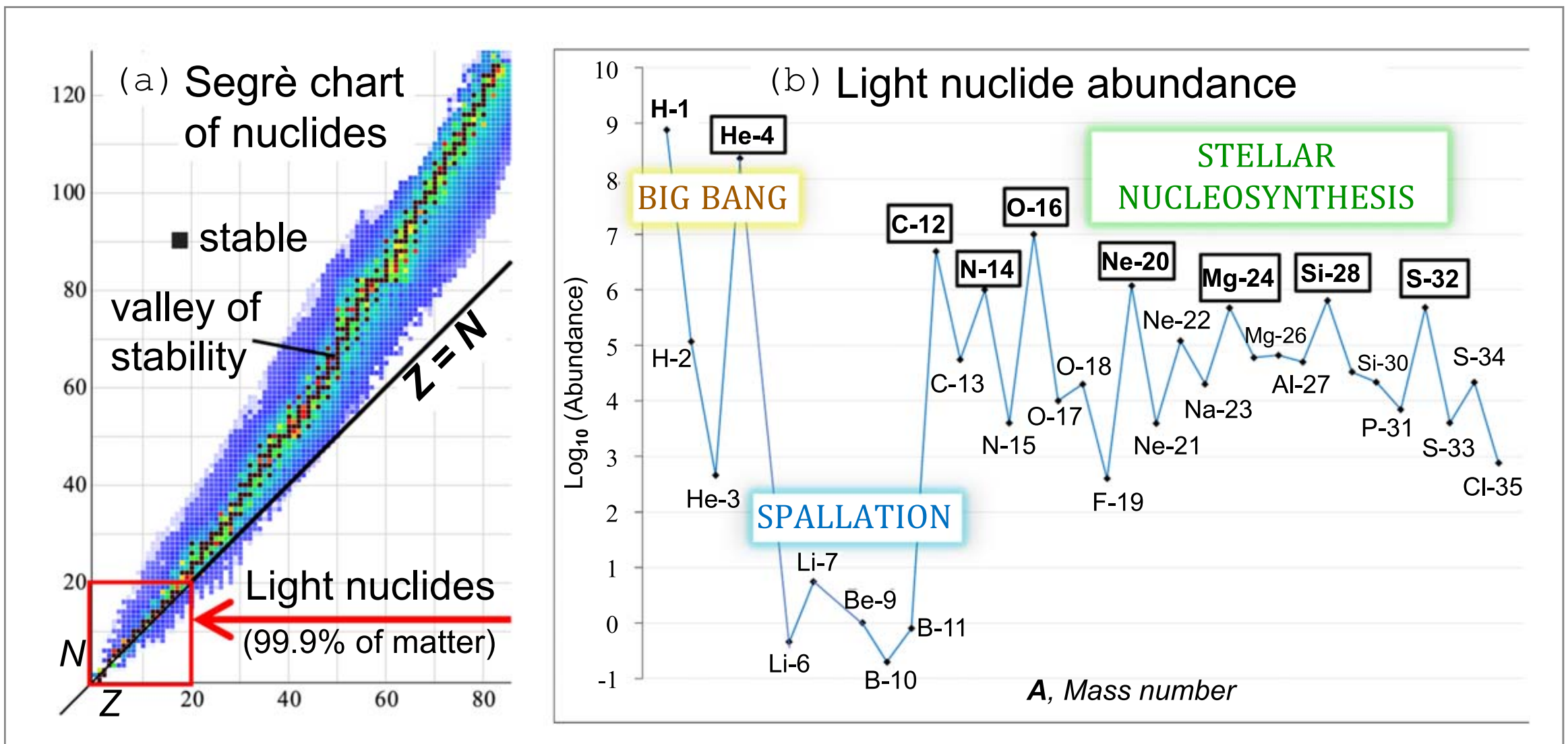

**Figure 3.** (a): The Segrè plot of the proton/neutron (Z/N) ratio of the ≈3300 known nuclides. The light nuclides comprise 99.9% of ordinary baryonic matter (red box). (b) The stable light nuclides through $^{35}$Cl plotted by mass number, and their nucleosynthetic origins. Aside from hydrogen, most ordinary baryonic matter in the universe (99.5%) can be found in just eight 'equinucleonic' nuclides (black boxes) containing equal numbers of protons and neutrons and thus fall on the Segrè Z = N line. The scarce spallation nuclides are too fragile to survive the highly kinetic reaction conditions of the Big Bang or stellar nucleosynthesis.

Nucleons pair up within nuclei to form short-range correlated pairs (SRCs) equivalent structurally to the deuteron [12]. These are predominantly proton-neutron pairs due to the action of the spin-dependent tensor part of the strong nuclear interaction [13]. More recent knock-out experiments using high-energy *electron* beams directed towards nuclei from carbon to lead demonstrate that *pn* SRC pairs are a universal phenomenon and similar in all nuclei [14]. The *pn* SRC pair represents a prototypic stable proton-neutron interaction, in contrast to the unstable proton–proton or neutron-neutron interactions.

The proton/neutron ratio of the most abundant stable nuclides also bears mentioning. Though hydrogen (the proton) comprises most of the ordinary matter of the universe (74%), when nucleons combine to form the most structurally resilient forms of polynucleonic matter, they tend to do so in a one-to-one ratio of protons to neutrons [15]. 99.5% of ordinary baryonic matter by mass (aside from hydrogen) contains equal numbers of protons and neutrons (figure 3), and it is these forms of polynucleonic matter that emerge in greatest abundance from stellar nucleosynthetic processes such as supernovae. That equality of nucleons correlates with relative cosmic superabundance may suggest that nucleon equality contributes to nuclide kinetic stability (resilient to destruction by kinetic impact forces) sufficient to survive the highly energetic nuclear reaction conditions of Big Bang and stellar nucleosynthesis.

In addition to structural and compositional factors, a more subtle but equally perplexing challenge relates to the sequence and occupancy of nucleons within nuclides through $^{36}$Ar. The list of stable nuclides progresses stepwise, one nucleon at a time, except for the unstable A = 5 and A = 8 amu nuclides. Why isobars of these two nuclides are unstable and why the sequence of stable nuclides progresses one nucleon at a time is unclear. Further confounding the matter, the selection of either a proton or a neutron at each successive step appears random. A complete structural model of the atomic nucleus ought to predict why adding one type of nucleon (and not the other) would result in the next stable nuclide while somehow accommodating the disruption in the stepwise sequence at the A = 5 and A = 8 nuclides.

Again, the compositional factor that most closely correlates with nuclide abundance is an equality of protons and neutrons. This factor is necessary, but not sufficient, for nuclide durability or kinetic resilience within energetic nuclear reaction conditions. The stable nuclides $^{6}$Li, $^{10}$B, and $^{36}$Ar, are among the rarest of naturally occurring nuclides [15], and the unstable nuclides $^{8}$Be, $^{18}$F, $^{22}$Na, $^{26}$Al, $^{30}$P, and $^{34}$Cl, also have equal numbers of protons and neutrons. Furthermore, $^{56}$Fe is one of the top 10 most abundant nuclides, but has an *unequal* number of nucleons (26 protons versus 30 neutrons). Its unusual abundance may arise from the formation of $^{56}$Fe as one of the last exothermic (thermodynamically favored) fusion reactions in the nucleosynthetic chain, resulting in the accumulation of $^{56}$Fe as an end product of supernovae nucleosynthesis [16].

While the lighter nuclides are generally more abundant than heavier nuclides (see figure 3), the elements lithium, beryllium, and boron are ≈ five orders of magnitude *less* abundant than the next three heavier elements, carbon, nitrogen, and oxygen (from which they derive) [15]. All six are initially produced by stellar nucleosynthesis, a highly kinetic environment that subjects nuclei to what structural engineers might term a 'dynamic impact load' (Salvadori 1980) [17]. However, the fragile isotopes of lithium, beryllium, and boron (along with $^3$He) are thought to be destroyed shortly after formation [18]. Instead, their natural occurrence is a result of spallation, a process involving the fission by cosmic rays of carbon, nitrogen, and oxygen, respectively [19].

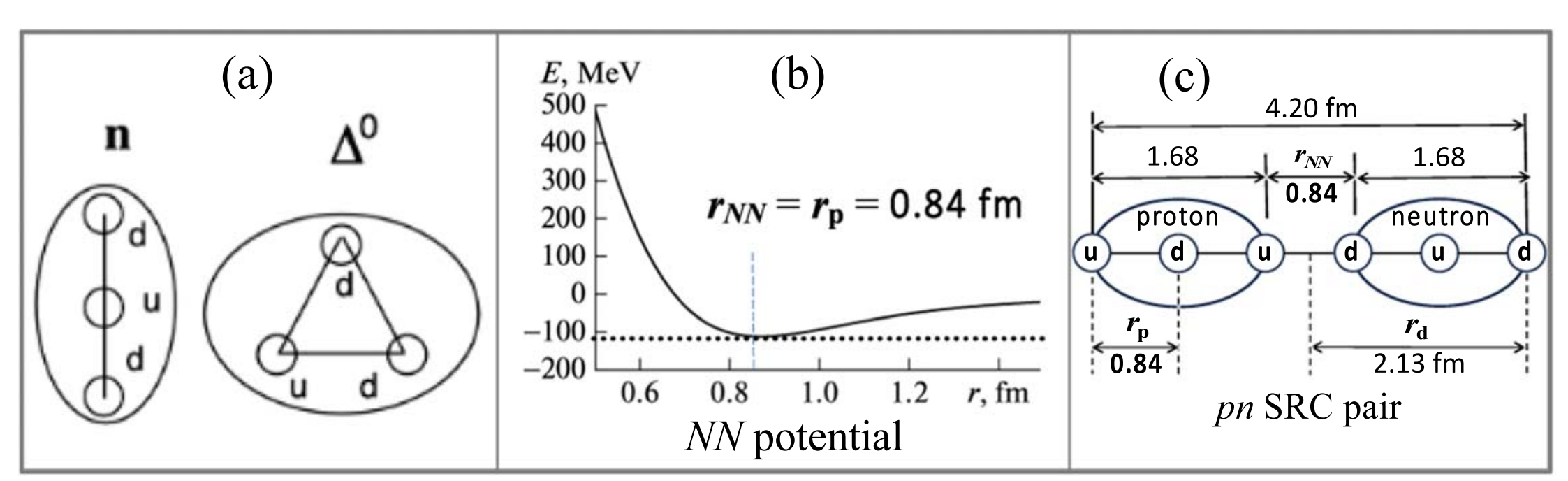

**Figure 4.** The qualitative quark picture (a) of the prolate neutron ground state n and oblate neutron excited state $\Delta^0$ 'Nucleon deformation and atomic spectroscopy,' Canadian Journal of Physics. 83(4): 455–465 [20], with permission from © Canadian Science Publishing. A similar linear ground state triplet may be obtained for the proton by exchanging u and d quarks. Panel 3(b) shows the nucleon-nucleon (*NN*) potential in the $^1S_0$ channel in the Argonne v18 potential is minimal at ≈0.8 fm [24], where the potential curve's slope = 0 (dotted line) and attractive forces between the two nucleons balance repulsive forces. Within the model, the *NN* separation distance $r_{NN}$ is set equal to the proton's charge radius $r_p = 0.84$ fm (superimposed by the author). Panel 3(c) shows the qualitative quark depiction of a prototypic proton-neutron short-range correlated pair (deuteron). The proton radius $r_p = 0.84$ fm is the distance from the proton's central down quark to an adjacent up quark, equivalent to the separation between nucleons set by the *NN* potential, $r_{NN} = 0.84$ fm. The deuteron's predicted radius of 2.10 fm (half the rotational diameter of 4.20 fm) correlates with the experimental deuteron charge radius $r_d = 2.13$ fm [25].

Here, we present a 3D coordinate-space model that is rigorously consistent with the accepted empirical knowledge of the atomic nucleus and yet remains accessible across all disciplines of science. Structures below $^7$Li derive from various regular polygon geometries. Above $^7$Li, nuclide structures build upon progressively larger anisotropic cylindrical lattices where one end of the nuclear cylinder exhibits steric features that determine the sequence and occupancy of successive nucleons, in effect a steric selection mechanism. This mechanism generally favors the formation of stable proton-neutron *NN* interactions (*pn* SRC pairs) while maintaining a stable nucleon ratio of neutrons ⩾ protons. Within the parameters of the model, bilateral nucleon symmetry correlates positively with cosmic abundance, while those isotopes in which symmetry is not possible (generally because proton and neutron numbers are unequal) correlate with scarcity.

Hypothetical interactions between sequences of regularly alternating fractional quark charges, modelling the proton, neutron, and deuteron as regularly alternating quark charge sequences, are shown below in potential-versus-distance plots characteristic of fusion potential curves. Within the model, the formation of a potential barrier is contingent upon the regular alternation of electromagnetic fields. At present there is no consensus that quark point charges regularly alternate in a predictable way that might generate the requisite fields. Thus, any conclusions drawn about alternating quark charges and a quark structure of the nucleus should be viewed with healthy scepticism. Nonetheless, the evidence for a prolate spheroid (cigar-shape) for the nucleon is well-established, and Buchman (2005) [20] has suggested a linear qualitative arrangement of alternating quarks to rationalize this shape [4]. Remarkably, should consensus ultimately coalesce around regularly alternating sequential quark geometry, the corresponding alternating electrostatic charge sequences of a pair of fusing deuterons could produce complex alternating/unequal electromagnetic fields sufficient to generate a potential energy barrier without recruiting the strong nuclear force. (Regardless of its capacity to generate a fusion potential, the electromagnetic fundamental force is not sufficient to hold nucleons together within the nucleus; For that, the strong nuclear force is required).

## 2. The alternating nucleon model

The shape of the hadron is relevant to the coordinate space packing of nucleons. Multiple studies of the intrinsic quadrupole moment of the proton have led to the expectation of a non-spherical nucleon shape. These studies analyze the transition to the first excited state of the nucleon, the Δ(1232) resonance [4, 21–23]. Figure 4(a) illustrates Buchman's 2005 rendition of the ground and excited states of the neutron [20]. The proton and neutron have nearly the same mass (938 MeV) and are otherwise identical in every respect except charge.

Buchman suggests that the repulsive spin–spin state between like-flavored quarks and the attractive spin–spin state between unlike-flavored quarks position the pair of like quarks at opposite ends of the prolate nucleon, leaving the unlike quark in the center of an overall linear arrangement of the quark triplet as depicted on the left of figure 4(a). This depiction is purely representational as individual quarks have never been isolated, and the uncertainty principle precludes knowing their exact locations and momenta. Additionally, the ground and excited state deformations appear to result from their respective quark distributions; in reality, each quark is presumed surrounded by a deformed cloud of quarks and antiquarks [23].

This qualitative quark triplet depiction proves useful in the geometric modeling of nuclide radii, regardless of its relationship to reality, in a manner perhaps analogous to the depiction of electron pairs in Lewis's dot structures. In the context of the model, qualitative quark depictions imply average quark positions that represent the spatial extent of the nucleon.

If atomic nuclei are bound systems of discreet protons and neutrons, a geometric representation of the nucleus must include consideration of the separation between them. The *NN* interaction determines this separation. It is derived phenomenologically and depends on factors such as spin, alignment, orbital motion, and local nuclear environment. Commonly used *NN* potentials include the Argonne AV18 potential, as well as the Paris, CD-Bonn, and Nijmegen potentials [26]. The force is the negative gradient of the potential energy, so attractive and repulsive forces balance at the bottom of the *NN* potential well in figure 4(b) where the slope = zero, defined by Ishkhanova at ≈0.8 fm [24]. For the purposes of predicting light nuclide radii, the best fit between model and experimental occurs when the approximation ≈0.8 fm is explicitly set equal to the proton's charge radius 0.84 fm [27]. The separation between nucleons derived from the *NN* potential, in concert with the qualitative quark triplet depiction, predict the deuteron's IAEA-accepted charge radius remarkably well, and extend generally to all stable nuclide radius predictions through $^{36}$Ar.

### 2.1. Big Bang nuclides

The deuteron is the prototypic representation of the ubiquitous short-ranged correlated pair, or *pn* SRC [14], a pair of strongly-bound nucleons whose separation distance is comparable to their radii. The geometric ball-and-stick representation of the deuteron's quark positions in figure 4(c) illustrates how the rotational diameter (4.20 fm) equals the sum of the proton and neutron lengths (1.68 fm each, twice the proton's radius) plus the separation distance between them derived from the *NN* potential (0.84 fm). This yields a rotational radius of 2.10 fm (half of 4.20 fm), a ≈99% fit to the experimental deuteron charge radius of 2.13 fm [25].

As shown in figure 4(c), the quark-to-quark spacing *between* nucleons is equivalent to the quark-to-quark spacing *within* nucleons, both equal to 0.84 fm. This leads to a generalizable regular 0.84 fm quark-to-quark spacing between sequential average quark positions without regard to nucleon positions. Within the model, the regular spacing exemplified by the deuteron is an essential constant, incorporated within the radius formula of a regular polygon (formula 1) to predict the charge radii of larger nuclides.

The list of stable nuclides through $^{36}$Ar progresses stepwise, one nucleon at a time, skipping over 5 and 8 amu nuclides. The next stable nuclide after the deuteron ($^{2}$H) is $^{3}$He, but the structure of $^{4}$He will be addressed first for the sake of expediency. One candidate structure for $^{4}$He is a qualitatively linear arrangement of its 12 quarks. The regular spacing of 0.8414 fm between sequential quarks yields a total length of 9.255 fm and a rotational radius half this at 4.628 fm. This is far greater than the experimental charge radius of 1.6755(28) fm [6] , making it necessary to consider a denser arrangement of nucleons. A circular arrangement provides the desired symmetry and density. A better correlation with the experimental charge radius of $^{4}$He arises from the qualitative positioning of its 12 quarks on the 12 vertices of a dodecagon (see $^{4}$He, figure 5). The predicted radius $r$ derives from the general radius formula of a regular polygon as shown in Formula (1), where $a$ is the distance between two neighboring quarks (0.8414 fm), and the number of sides (or vertices) $n$ equals 12, the number of quarks:

$$r = \frac{a}{2}\csc\left[\frac{\pi}{n}\right] \tag{1}$$

The regular polygon formula predicts a $^{4}$He radius of 1.625 fm, which falls within the range values included in Angeli(1999) [28] (1.61–1.70, n = 11(*e*-scattering)) and is a close approximation to the IAEA-accepted charge radius of 1.6755(28) fm [6]. Importantly, this point-symmetrical circular isomer is consistent with the empirical knowledge of helium structure, which indicates a hole or central depression in the electron scattering radial charge densities of $^{3}$He and $^{4}$He (left, figure 2), as reported by McCarthy *et al* [9]. (Many of the remaining light nuclides though $^{36}$Ar also demonstrate a similar central depression in the charge densities [11].)

There is a precipitous drop in abundance after the helium isotopes in the plot of cosmic abundance as there is no stable A = 5 nuclide. The half-lives of the isobars $^{5}$He and $^{5}$Li are measured in yoctoseconds (one preceded by a decimal point and 22 zeros), and this mass gap effectively halts Big Bang proton–proton nucleosynthesis from continuing beyond $^{4}$He [29]. The alternating nucleon model predicts this mass gap as no A = 5 nuclide can form an uninterrupted ring of alternating nucleons analogous to the ringed arrangement of $^{4}$He. The proton/neutron bound state is the only stable *NN* bound state, and at some point, in a closed circular arrangement of five nucleons, there would be an unstable proton–proton or neutron-neutron interaction precluding ring closure. The alternating nucleon model thus predicts that no 5 amu nuclide is stable.

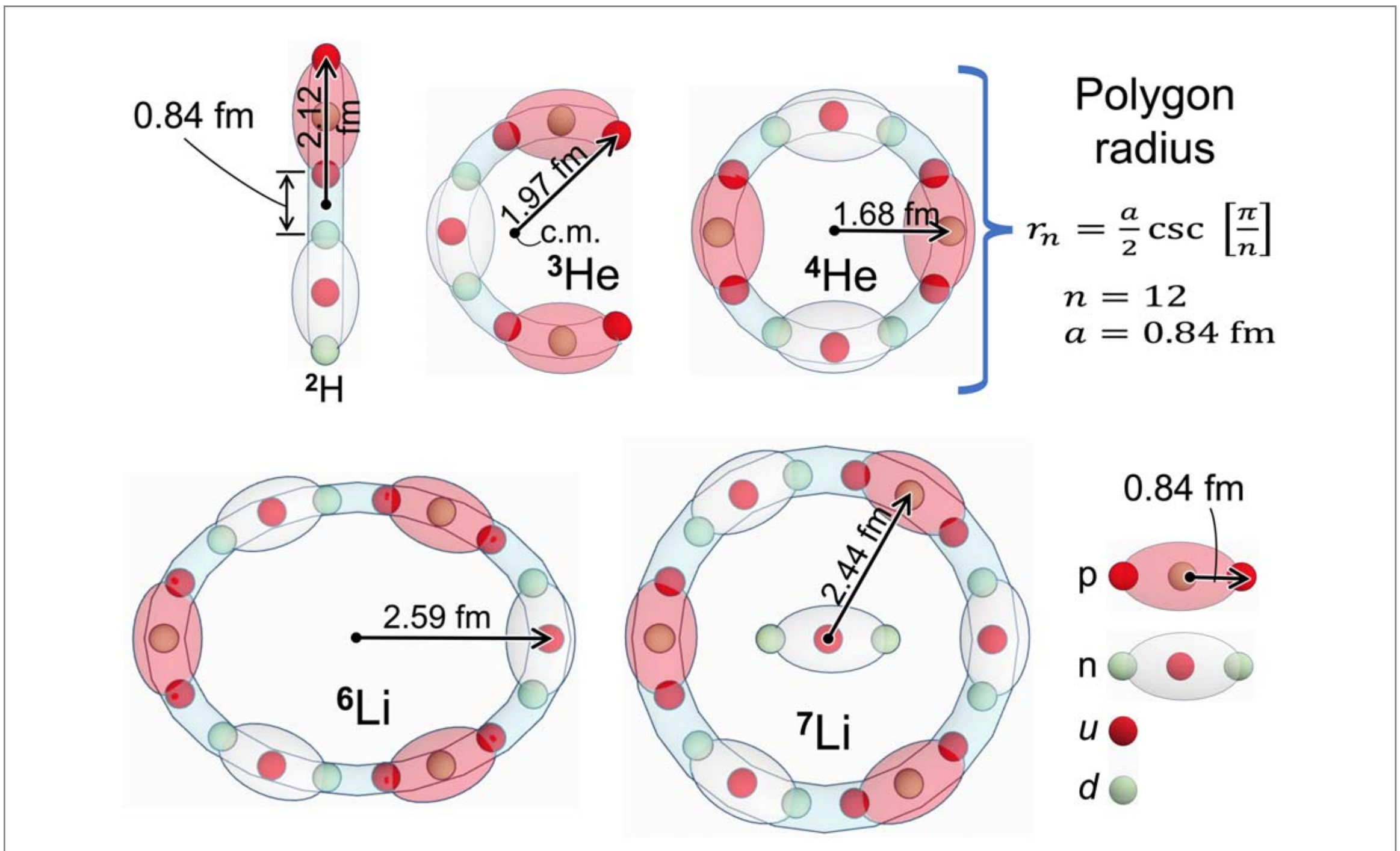

**Figure 5.** Polygon-derived geometries of the lightest stable nuclides based on the qualitative quark positions in figure 4(a) produce the erratic radius-to-mass plot below $^7$Li in figure 1. $^3$He and $^4$He derive from a regular 12-gon, while $^6$Li and $^7$Li derive from a regular 18-gon. Nucleon alternation intrinsically produces equal numbers of protons and neutrons in light nuclides. The separation between sequential nucleons is equivalent to the proton's radius. Rotation may distort $^6$Li's regular polygon shape into an oval with the long-axis radius as shown. Nucleons and nuclides are drawn to scale relative to the accepted charge radius values.

### 2.2. Spallation nuclides

The best-fit structure of $^3$He derives from the same regular dodecagon geometry employed to predict the radius of $^4$He as described above. $^3$He has only nine quarks, however, so three of the 12 dodecagon vertices are unfilled, and the new center of mass calculated by weighted average. Based on the distance from the center of mass to the farthest quark (see $^3$He, figure 5), the predicted radius of rotation is 1.92 fm (versus the accepted charge radius of 1.9661(30) fm [1]), well within the range of the $^3$He charge radius values included in Angeli (1999) [28] (1.84-1.97 fm, n = 9 (*e*-scattering)).

The radius prediction of $^6$Li derives from Formula 1 and a qualitative distribution of its 18 quarks into a regular octadecagon (18-gon). The predicted radius of 2.423 fm is within the range of reported charge radius values included in Angeli (1999) [28] (2.38-2.76 fm, n = 7 (*e*-scattering)). If the ring arrangement of *four* nucleons in $^4$He produces a kinetically stable structure (as indicated by its abundance and low $\Delta_{tot}R = 0.0028$ fm [6]) then a ring arrangement of six nucleons may result in a more lax structure for $^6$Li. Ring laxity and nuclear rotation may then distort the octadecagon into an oval (see $^6$Li, figure 5) with a long-axis radius of rotation closer to the IAEA accepted charge radius of 2.5890(390) fm. [6] Alternatively, as with any lax ring, the structure may demonstrate stable node/anti-node oscillation states [30], in which the long axis of the oval oscillates between 0° and ≈90°. Furthermore, the proposed lax or floppy structure of $^6$Li may account for its unusually large uncertainty of charge radius measurement ($\Delta_{tot}R = 0.0390$ fm [6]) across an unusually wide range of reported experimental radius values (2.38-2.76 fm) by electron (*e*)-scattering [28]. The significantly larger charge radius values obtained by muon ($\mu$)-scattering (3.1–4.2 fm [28]) may relate to the greater mass of the muon (200x greater than the electron), which may mechanically perturb or distort the $^6$Li structure during $\mu$-scattering experiments, resulting in a larger range of muon charge radii.

$^7$Li is a $^6$Li ring with a neutron inclusion in the ring's center (see $^7$Li, figure 5). The neutron inclusion is similar in principle to the orbiting nucleons of a halo nucleus [31]. Rather than orbiting around the nucleus in a halo, the best-fit radius prediction for $^7$Li includes the neutron within the interior of the loose oval of $^6$Li, essentially inflating or transforming the oval into a ring. As with $^6$Li above, the radius is predicted by formula 1 and a regular octadecagon arrangement of 18 quarks. The resulting 2.423 fm prediction is a near-perfect match to the experimental 2.4440(0.0420) fm [6], and well within the range of the $^7$Li charge radius values included in Angeli (1999) [28] (2.38–2.44 fm, n = 4 (*e*-scattering)). $^7$Li is an order of magnitude *more* abundant than $^6$Li,

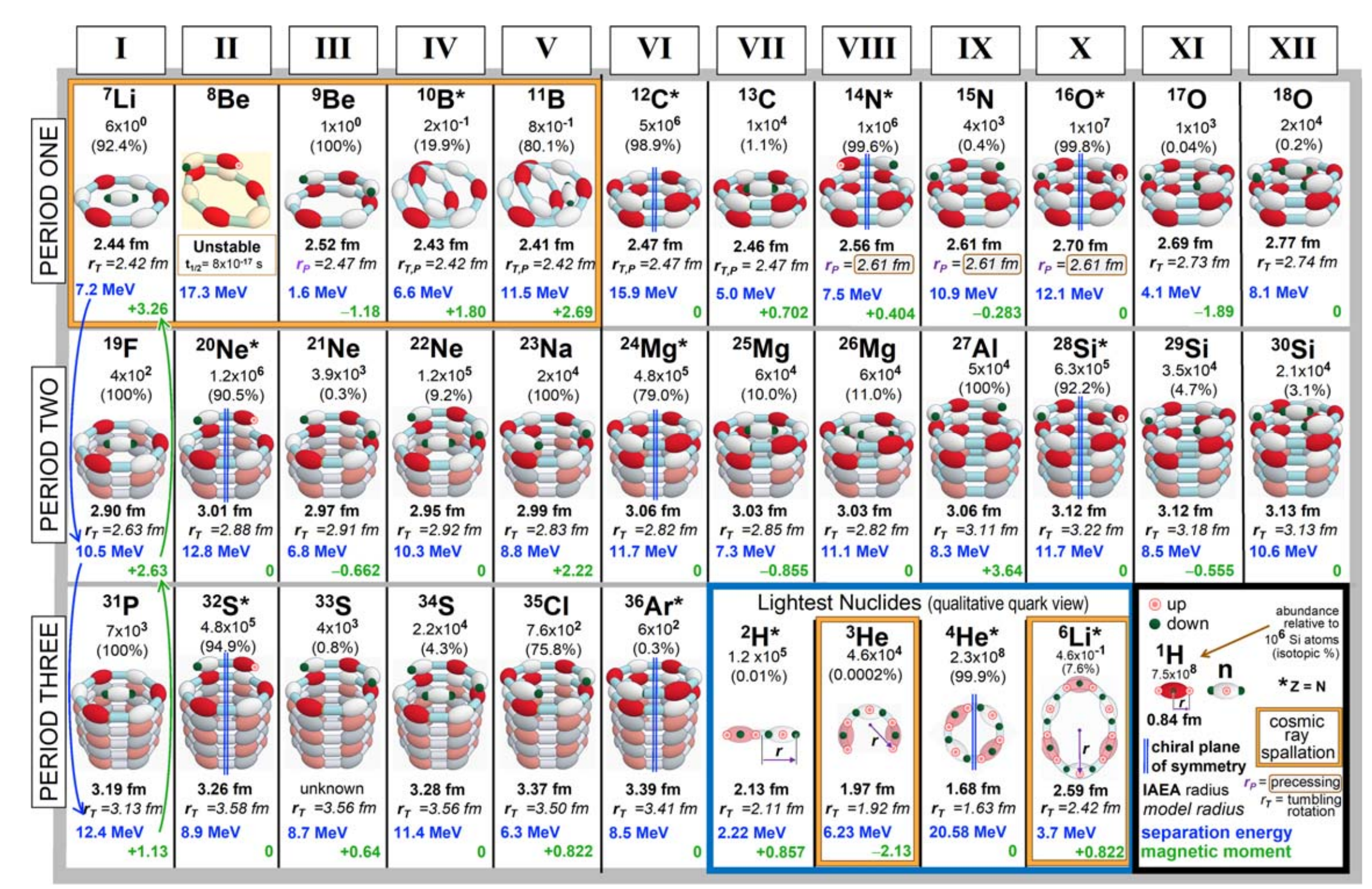

**Figure 6.** Alternating nucleon model structures of the stable nuclides through $^{36}$Ar. Columns I-XII represent unique nucleon configurations called anisotropes that repeat in Periods Two and Three and correspond to loose trends in nucleon separation energy and magnetic moments. Symmetrical structures designated with a blue double line correspond to the most abundant nuclides.

which may imply greater kinetic stability. The neutron inclusion within $^{7}$Li may serve as an architectural cross-member that imparts greater kinetic stability, stabilizing the lax $^{6}$Li ring by distributing dynamic impact loads, similar in principle to the function of a cross-member within a bridge truss [32].

$^{8}$Be is an alpha nuclide with a half-life $\approx 10^{-16}$ s. The $^{8}$Be mass and instability suggest a transient bound state of two helium nuclei. The present representational model predicts a nucleon sequence in which a proton added to the $^{7}$Li neutron inclusion would create a stable *pn* SRC pair, and that the SRC would then associate with the $^{6}$Li ring within a $^{6}$Li + $^{2}$H bound state. The proposed structure of the unstable nuclide is shown in figure 6 below. However, given that there is no stable A = 8 nuclide, it may be that the proposed node/anti-node vibrational instability of the lax $^{6}$Li ring structure (as discussed above) might be sufficiently destabilizing to preclude stable binding to $^{2}$H.

The best-fit alternating nucleon structure of $^{9}$Be combines a $^{6}$Li base ring bound to a 3 nucleon *npn* sequence on a second incomplete ring, as shown in figure 7. The distance between the $^{6}$Li base ring and 3 nucleon *npn* sequence is 0.9616 fm (determined from the structure of $^{12}$C below). The predicted radius of 2.47 fm is well within the range of the $^{9}$Be charge radius values included in Angeli (1999) [28] (2.20-2.53 fm, n = 9 (*e*-scattering)), and a close match to the IAEA-accepted value of 2.5190(0.0120) fm (Angeli *et al* 2013) [6]. Importantly, each of the three nucleons on the second ring is associated with its opposite isospin nucleon on the $^{6}$Li base ring to form proton-neutron pairing of another type, as shown in figure 7.

The next two stable nuclides, $^{10}$B and $^{11}$B, have roughly the same experimental charge radius as $^{7}$Li. All three radii $\approx$2.42 fm, essentially the same radius (2.43 fm) predicted for a regular octadecagon by formula (1) where $n = 18$ quarks ($^{6}$Li) and $a$ is the proton's radius (0.8414 fm). Within the model, $^{7}$Li, $^{10}$B, and $^{11}$B all have a base $^{6}$Li ring structure with the mass of their additional nucleons included within the interior of the $^{6}$Li ring. Thus, the predicted radii of all three nuclides equate to the radius of the exterior base $^{6}$Li ring. $^{7}$Li has a neutron inclusion, as discussed above, but within the proposed $^{10}$B structure, the exterior $^{6}$Li ring loosely includes a ringed four-nucleon $^{4}$He nucleus instead of a neutron. The extremely low cosmic abundance of $^{10}$B is consistent with a loose association between the rings. The proposed structure of $^{11}$B is the same as $^{10}$B but with an additional neutron tethering the base $^{6}$Li ring to the included $^{4}$He ring (figure 6), which may have a stabilizing effect as reflected in the 4x greater abundance of $^{11}$B compared to $^{10}$B (see figure 3(b)).

### 2.3. $^{12}$C-based nuclides

In mechanical systems, the lamination of flexible materials confers both stiffness and strength [33]. The proposed structure of $^{12}$C comprises two stacked $^{6}$Li rings (figure 8). The open ring structure is consistent with the central dip in the experimental $^{12}$C point-proton density (Carlson *et al* 2015) [34]. Although an individual $^{6}$Li ring is lax (as discussed previously), and the low abundance may indicate structural fragility, a bound state of two stacked $^{6}$Li rings effectively laminated together by the strong force within $^{12}$C would constitute a more rigid and durable structure, thus accounting for the unusually high cosmic abundance of $^{12}$C.

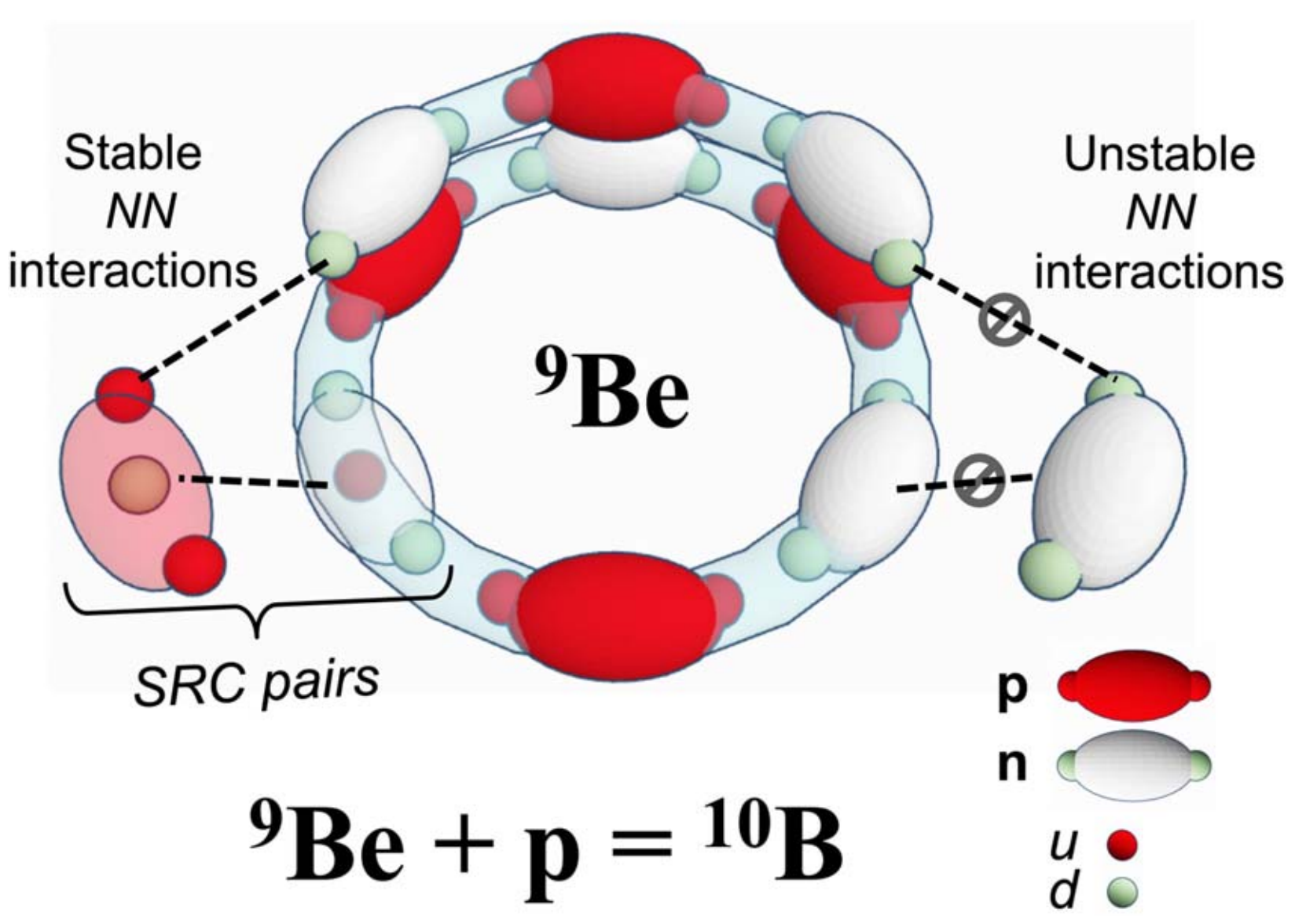

**Figure 7.** The addition of a proton to $^{9}$Be (panel left) creates two stable nucleon-nucleon interactions (two *pn* SRC pairs). Conversely, the addition of a neutron to $^{9}$Be (panel right) creates unstable nucleon-nucleon interactions. The structure of $^{9}$Be thus sterically predicts that the addition of a proton will produce the next stable nuclide, in this case $^{10}$B, while the addition of a neutron makes unstable $^{10}$Be ($t_{1/2} = 1.4 \times 10^{6}$ yrs).

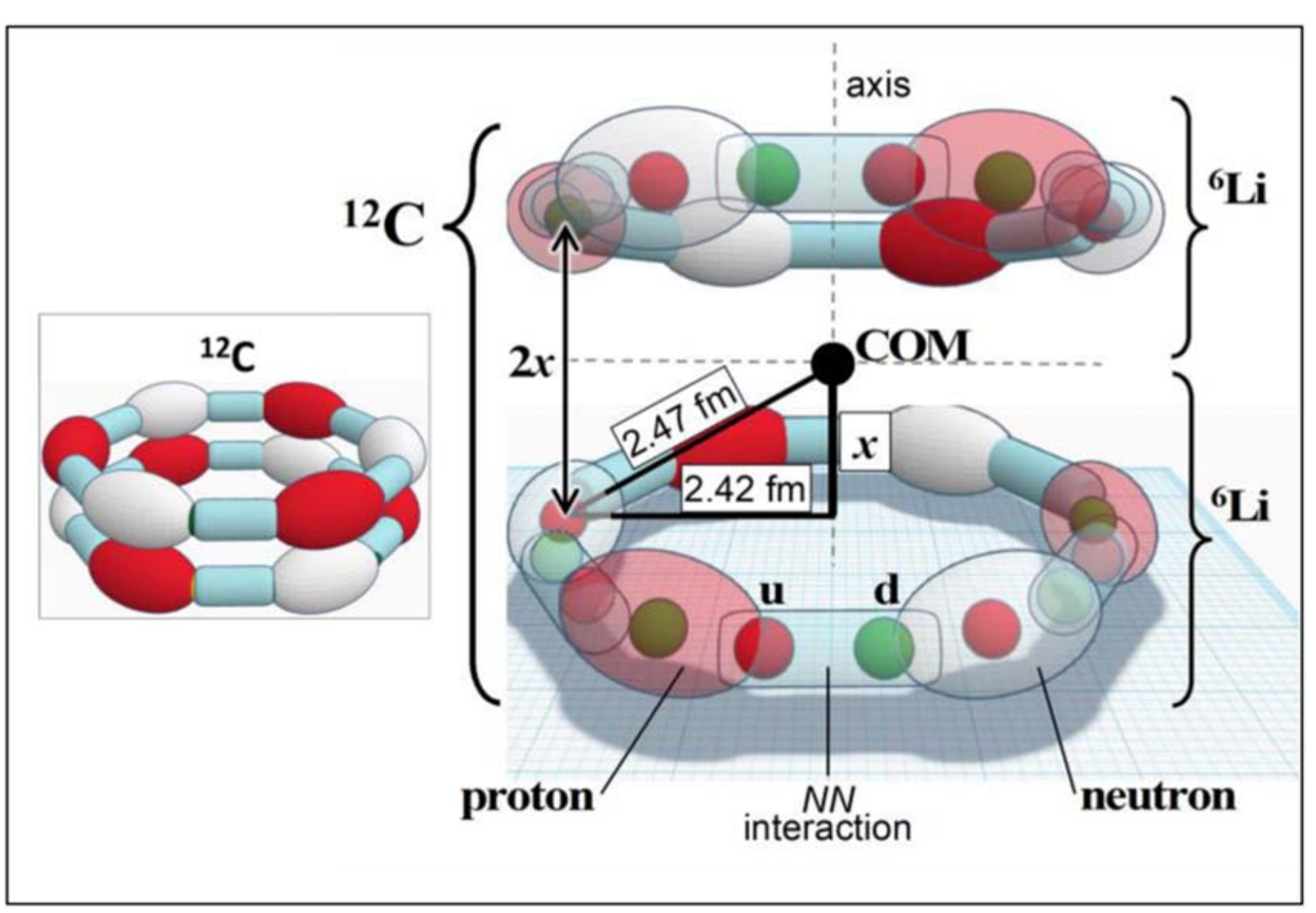

**Figure 8.** An expanded view of $^{12}$C illustrates a geometric method of determining the distance between stacked rings in larger nuclides. The distance from the center of mass (COM) to any quark position is 2.4702 fm (the $^{12}$C charge radius). Likewise, the distance from the geometric center of each $^{6}$Li ring to any of its 18 average quark positions equals 2.423 (derived from Formula 1). Solving for x using the Pythagorean theorem yields 0.4808 fm, and twice this is 0.9616 fm, the model's standard separation between stacked 6-nucleon rings.

If $^{12}$C is a pair of stacked $^{6}$Li rings then each 6-nucleon rings would distribute 18 quarks (6 nucleons * 3 quarks/nucleon) on the vertices of a regular octadecagon having sides $a = 0.8414$ fm and $r = 2.423$ fm per Formula 1. The proposed structure of $^{12}$C exhibits perfect point symmetry. This implies that the distance from the $^{12}$C center of mass to any of its 36 qualitative quark positions is the same and equal to the IAEA-accepted charge radius of 2.4702 fm. The $^{12}$C abundance, symmetry, and small uncertainty in measurement ($\Delta$tot R = 0.0022 fm [6]) are consistent with stability and rigidity, making $^{12}$C ideal for setting the standard distance between adjacent parallel 6-nucleon rings. Like the separation between nucleons derived from the *NN* potential, the separation between 6-nucleon rings is necessary for predicting the radii of the majority of light nuclides above $^{12}$C in figure 6. The charge radius and perfect point symmetry of $^{12}$C geometrically fixes the ring separation between stacked 6-nucleon rings at 0.9616 fm as detailed in figure 8. (As an aside, if the proton's radius 0.8414 fm is equivalent to the long-axis radius of a prolate spheroid, then the short-axis radius is equivalent to $x = 0.4808$ fm, as shown in figure 8).

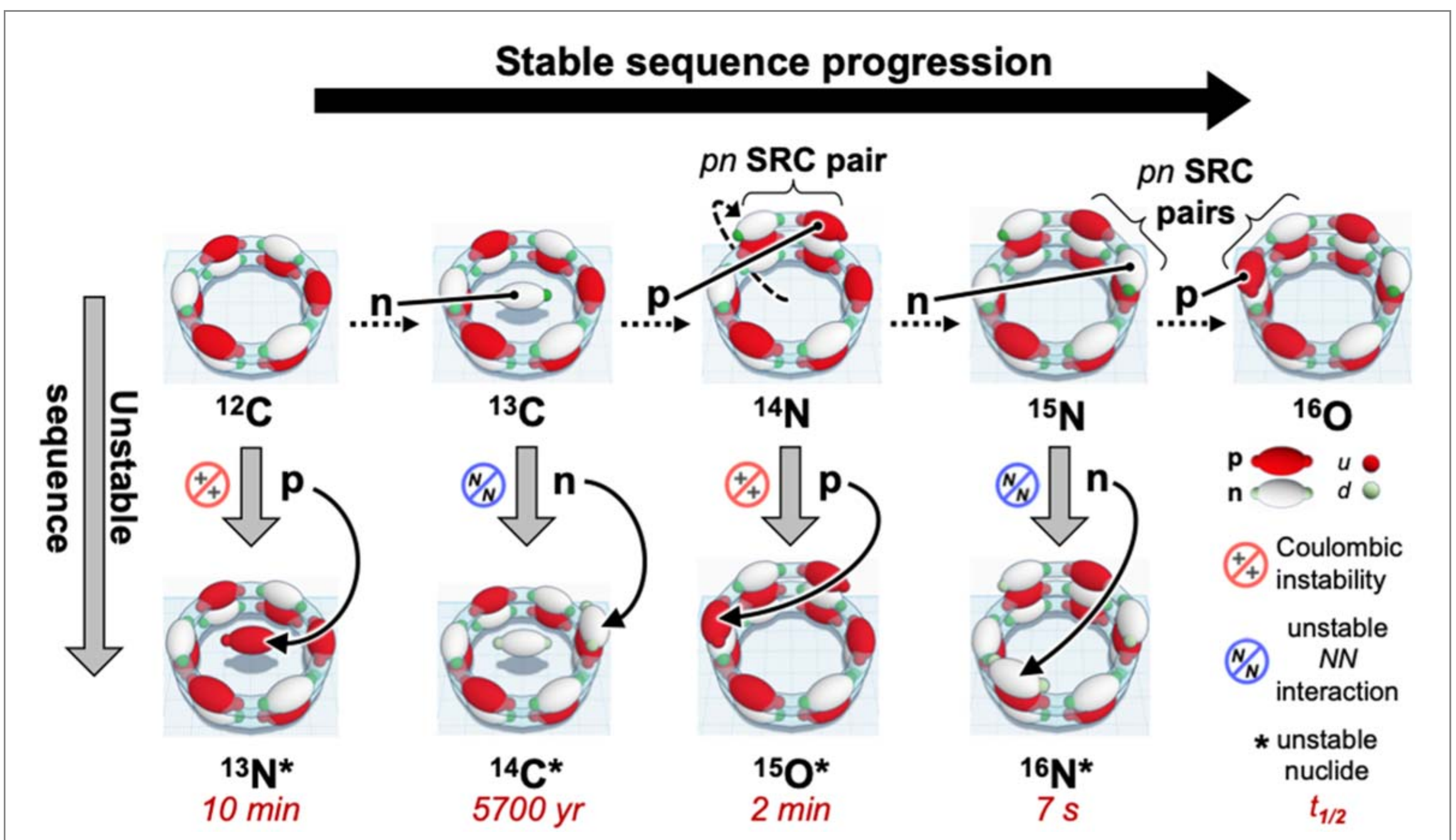

**Figure 9.** The nucleon structure of a light nuclide sterically predicts whether an additional proton or neutron will be incorporated within the next stable nuclide. The sequence generally favors the formation of one or more stable NN interactions (i.e., pn SRC pairs). If the addition of either nucleon (as in the case of $^{14}$N), or neither nucleon (as in the case of $^{12}$C), results in a stable sequential NN interaction, the progression defaults to adding a neutron in order to maintain a stable ratio of neutrons $\geqslant$ protons. (This flow diagram intends to indicate a sequence and occupancy of nucleons rather than the sequence of nucleosynthesis.).

$^{12}$C and $^{13}$C have nearly identical radii (2.47 fm and 2.46 fm, respectively), even though $^{13}$C has an additional neutron (figure 9). This is similar to the cases of $^{7}$Li, $^{10}$B, and $^{11}$B discussed above, all three of which have the same charge radius and additional mass included within a base $^{6}$Li ring. Here, the additional neutron of $^{13}$C is included within the $^{12}$C structure and, therefore, has a charge radius essentially identical to $^{12}$C. The relative abundance of $^{13}$C (two orders of magnitude less than $^{12}$C) suggests that the additional neutron may *not* confer structural durability, unlike the neutron inclusions of $^{7}$Li and $^{11}$B.

While a *neutron* added to $^{12}$C results in stable $^{13}$C, a *proton* added to $^{12}$C is disfavored as unstable $^{13}$N (figure 9) has an unfavorable ratio of protons > neutrons. As the sequence progresses, adding a proton to $^{13}$C to form $^{14}$N results in greater stability as this additional proton links to the central neutron to form a *pn* SRC pair. Conversely, the addition of a neutron to $^{13}$C results in an unstable neutron-neutron interaction forming unstable, radioactive $^{14}$C (figure 9). After calculating the center of mass (COM) and the distance from the COM to the farthest qualitative quark position, the best fit to the experimental radius of $^{14}$N places the *pn* SRC on a third parallel partial ring (figure 9). The model radius prediction of 2.61 fm closely matches the value accepted by the IAEA of 2.5582(0.0070) fm [6]. (Note: The progressive addition of individual nucleons is *not* intended to indicate nucleosynthetic reactions but rather to indicate a steric role for the sequence and occupancy of nucleons.)

As the sequence progresses, the addition of a neutron to the third parallel partial ring of $^{14}$N creates a second *pn* SRC pair with the proton already there, yielding $^{15}$N whose predicted radius of 2.61 fm matches the value accepted by the IAEA of 2.6058 (0.0080) fm. Importantly, the addition of a proton to $^{14}$N would *also* create a *pn* SRC pair, but forms unstable $^{15}$O (figure 9). In this case, the addition of the proton is disfavored due to the resultant unstable ratio of protons > neutrons.

The model structure of $^{15}$N has a 3-nucleon *npn* sequence on a third incomplete ring bound to a $^{12}$C. The addition of a proton to either of the two neutrons already present on the third ring enhances overall stability by creating a pn SRC pair to form stable $^{16}$O, with a predicted radius of 2.61fm as compared to an experimental radius of 2.6991(0.0052) fm [6]. Conversely, there is no placement of an additional *neutron* that would create a *pn* SRC pair. *The steric structure of* $^{15}$*N thus selects which type of nucleon may be added to form the next stable nuclide.* Figure 9 generally illustrates how steric factors guide nucleon selection in predicting nuclide stability.

The model structure of $^{16}$O contains a complete $^{12}$C double ring plus a third partially-filled ring containing a sequence of four alternating nucleons beginning with a neutron and ending with a proton (figure 9, upper right). The addition of either a proton to one end or a neutron to the other could create a stable *pn* SRC pair. However, an additional proton would result in an unstable ratio of protons > neutrons, so the addition of a neutron is favored. The next stable nuclide is thus $^{17}$O with a predicted radius of 2.73 fm, compared to an experimental value of 2.6932(0.0075) fm [6].

The structure of $^{17}$O contains a $^{12}$C double ring and a sequence of five alternating nucleons. The sequence begins and ends with a neutron. Per the aforementioned selection rules, the addition of a proton would bind to both these neutrons, close the ring, and create two new *pn* SRC pairs. However, the resulting $^{18}$F is unstable. The predicted structure has a $^{6}$Li ring stacked on a $^{12}$C ring, and the model predicts that its instability may relate to the node/antidote vibration of $^{6}$Li discussed above, precluding an effective interaction with $^{12}$C. The next stable nuclide in the sequence instead is $^{18}$O, whose structure is identical to $^{17}$O but with an additional neutron included inside the nuclear cylinder. The predicted radius of $^{18}$O is 2.74 fm, while the IAEA-accepted radius is 2.7726(0.0056) fm. The low relative cosmic abundance of $^{18}$O ($2 \times 10^4$) suggests it may be the product of a minor competing nuclear reaction while the major reaction produces $^{18}$F, which is unstable and immediately decays.

The remaining nuclides through $^{36}$Ar add one nucleon at a time according to the steric selection method illustrated in figure 9. Predicted versus experimental charge radii are included in figure 6.

### 2.4. Rotational considerations in radius prediction.

The model assumes that a nucleus rotates in space and that the manner of rotation determines its spatial extent and, therefore, its best-fit radius. Any rotation in a three-dimensional space can be analysed using a combination of the three principal axes, and where a rigid body has an axis of symmetry, this axis will correspond to a principal axis of rotation [35].The most stable rotational axis is where the greatest mass is distributed the farthest from the rotational axis, and rotation about this major inertia axis corresponds to the minimal kinetic energy.

The radii of most light nuclides are best predicted by end-over-end tumbling rotation (indicated by italics that are preceded by $r_T$ in figure 6), in which case the radius is determined by the distance from its center of mass to its farthest average quark position. This implies a radius of rotation through the nucleus COM and orthogonal to the length of the model nuclear cylinder. In some of the lightest nuclides, rotation may look more like the precession of a plate spinning on a table just before it comes to rest. In this case, the method of radius determination assumes the rotational radius is half the distance between the two average quark positions farthest from one another within the proposed nuclide structure. The rotational axis, in this case, passes through the geometric center of the base $^{6}$Li ring and is orthogonal to the geometric plane occupied by the $^{6}$Li ring. For many of the nuclides $^{7}$Li through $^{16}$O, the manner of rotation made no difference to the radius prediction. However, in the cases of $^{14}$N (experimental charge radius = 2.56 fm), $^{15}$N (2.61 fm), and $^{16}$O (2.70 fm), the precessing radius ($r_P$) determinations of 2.61, 2.61, and 2.61 fm (respectively) were a better fit to experimental than the tumbling radius predictions of 2.79, 2.84, and 2.80 fm (respectively). The experimental charge radius of $^{16}$O (2.70 fm) is halfway between the precessing prediction of 2.61 fm and the tumbling prediction of 2.80 fm and may represent a mix of the two rotational states. As an aside, the manner of rotation may be a factor in a nuclide's nuclear magnetic moment.

## 3. Results

Figure 6 illustrates the alternating nucleon model structures of the light nuclides through $^{36}$Ar. The twelve Roman numerals across the top of figure 6 refer to the 'anisotrope' group number, defined as the number of nucleons forming the anisotropic end of the cylindrical structure of each nuclide. The anisotrope serves as a sort of steric active site in guiding the sequence and occupancy of nucleons, while the other end of the cylindrical structure remains functionally inert. Best-fit radius predictions result when nucleons are added *only* to the anisotrope end of the evolving nuclear cylinder, as described in the Alternating Nucleon Model section. The anisotrope is depicted in red and white nucleons, while the essentially inert remainder is depicted in gray and pink nucleons. Except for the boron nuclides, the anisotrope group number also corresponds to the unique nucleon configuration of the evolving anisotropic end of each structure.

There are three Periods listed down the left side of figure 6. The structure of the anisotrope generally evolves in the same way across each Period by adding one nucleon at a time in a predictable configuration. The boron isotopes $^{10}$B and $^{11}$B represent an exception; however, their smaller-than-expected experimental radii present a challenge to structural elucidation, which is resolved by placing a $^{4}$He ring nucleus within the center of each, as shown. The depiction of $^{10}$B in figure 9 as a four-nucleon sequence atop a 6-nucleon ($^{6}$Li) ring may represent an intermediary state, and the oscillation of the floppy $^{6}$Li base may facilitate the joining of the ends of the four-nucleon sequence to form an additional *pn* SRC and complete the $^{4}$He ring, which then settles precariously within the center of the $^{6}$Li ring as an inclusion. This structural contortion and potential weakness in the structure may account for boron's low cosmic abundance (discussed previously) [36]. As an aside, the inclusion of neutrons and alpha particles within the center of the boron nuclides may set a precedent for the increased density of nuclide structures above $^{36}$Ar.

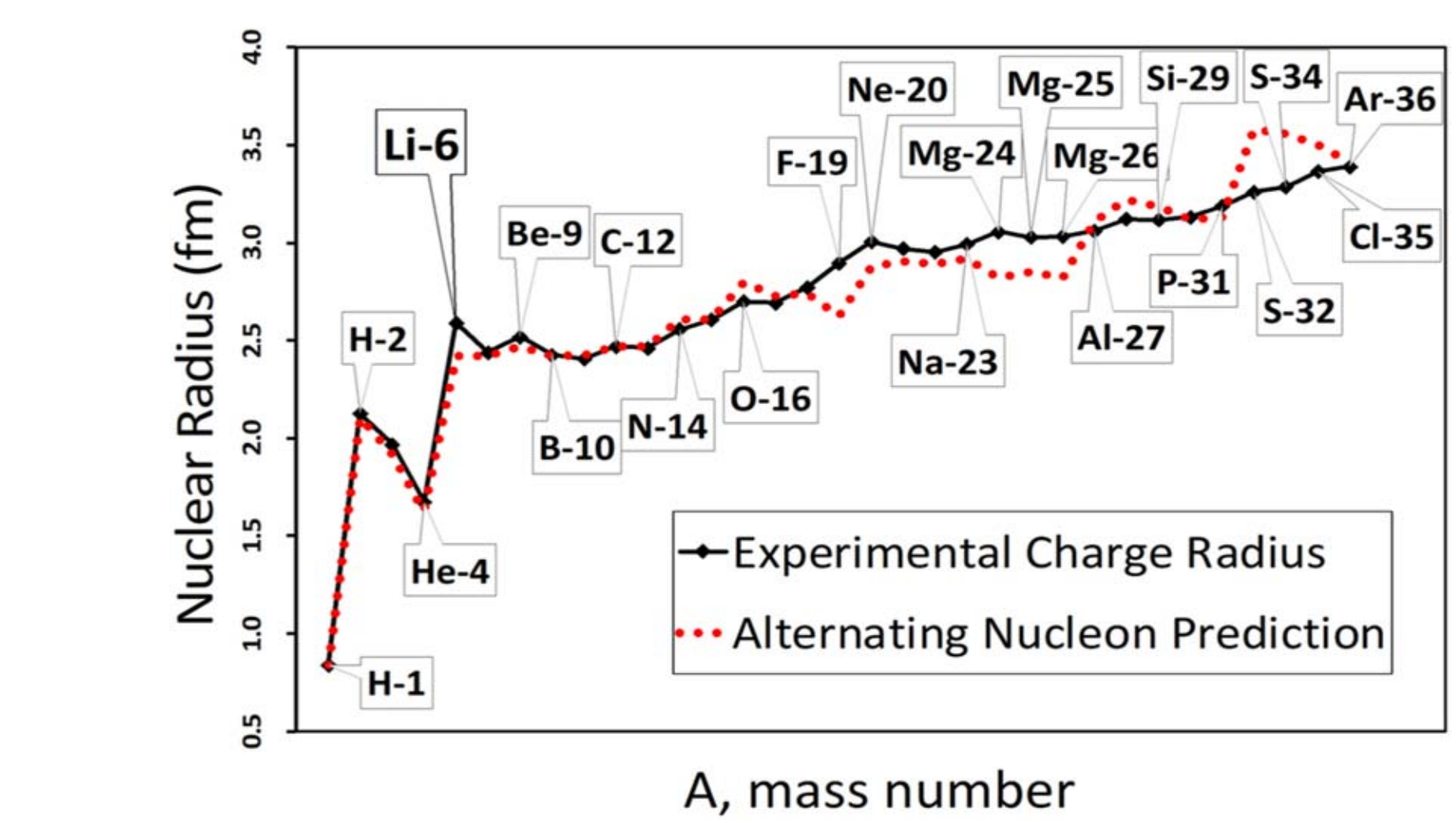

**Figure 10.** Model charge radius predictions demonstrate strong positive correlation with IAEA-accepted charge radii (black solid line), (r(31) =.98, p<.001). Polygonal geometries contribute to the erratic radius-to-mass below Li-6 while the incremental addition of one nucleon at a time to the model Li-6 ring accounts for the linearity above Li-6.

The tabulation of nucleon structures in figure 6 begins with $^{7}$Li, corresponding to the linear portion of the radius-to-mass plot depicted in figure 1. Each nuclide is enclosed in a box and the legend is shown in the table's lower right corner. Below each nuclide symbol is the cosmic abundance cited as the number of atoms relative to silicon, a standard measure set arbitrarily at one million ($1 \times 10^{6}$). The stable isotopic percentage within an element is shown in parenthesis below this; for elements that have only one stable isotope, such as $^{23}$Na, the number is 100%. Below the graphic depiction of each cylindrical nuclide is the IAEA-accepted charge radius in bold, followed by the model-predicted radius in italics. As shown in figure 10, predicted radii track well with experimental, demonstrate strong positive statistical correlation with experimental ($r(31) = .98, p < .001$). The manner of rotation affects both the radius prediction and the nuclear magnetic moment (discussed below), and can be either end-over-end tumbling ($r_T$) or precessing ($r_P$), as discussed previously.

Below the radius prediction is the nucleon separation energy (determined from the difference in total binding energy between the nuclide of interest and the previous stable nuclide, generally 1 amu lighter). The nucleon separation energy trends are confined to a group (column), either increases or decreases down the group. The nuclides within each group contain the same anisotrope, and trends within a group may be related to the structure and composition of the anisotrope. The trend is downward in group I, and upward in group II. Thereafter, the trends alternate regularly by two so that the trend in groups III and IV is up, groups V and VI down, groups VII and VIII up, groups IX and X down, and finally, groups XI and XII up. The cause of this pattern warrants further investigation.

The bottom number in each nuclide's box is the nuclear magnetic moment ($\mu$), an experimentally determined value derived from an intrinsic and orbital angular momentum, the latter relating to how a nucleus rotates in space [37–39]. Manner of rotation is particularly relevant when considering that the best-fit manner of rotation of most nuclides in figure 6 is a tumbling rotation while for others it is precessing, and for some it can be a mix of the two. Due to nucleon pairing effects, $\mu = 0$ when the numbers of protons and neutrons are both even. Since even group numbers have even numbers of protons and neutrons, their magnetic moments will generally equal to zero, as shown in figure 6. Otherwise, the nuclear magnetic moment $\mu$ is unique for each nuclide and varies in sign and magnitude as mass number increases. The nuclear magnetic moments within each group tend to maintain the same sign while decreasing in magnitude with increasing period number, with the exception of groups VII and IX, in which the sign differs between periods. The sign disparity in groups VII and IX between Periods I and II may relate to the effects of precessing rotation on the magnetic moments of the nuclides in Period One.

The eight equinucleonic (Z = N) nuclides $^{4}_{2}$He, $^{12}_{6}$C, $^{14}_{7}$N, $^{16}_{8}$O, $^{20}_{10}$Ne, $^{24}_{12}$Mg, $^{28}_{14}$Si, and $^{32}_{16}$S, (marked with asterisks *) exhibit a chiral bilateral plane of symmetry (double blue line). Together the Big Eight nuclides alone comprise 99.5% of all polynucleonic baryonic matter. Above $^{35}$Cl, however, structural symmetry is no longer a sensitive and specific predictor of abundance. Within the model, $^{36}$Ar is symmetrical but comprises only 0.3% of elemental argon, while $^{40}$Ar is argon's most abundant isotope at 99.66%. Neither do the equal numbers of protons and neutrons in $^{6}$Li result in a rigid, symmetrical nucleus through which a plane of symmetry may be drawn. The $^{4}$He ring inclusion of $^{10}$B confers asymmetry despite its equal proton/neutron numbers (as shown), precluding a plane of symmetry. Nuclei containing a neutron inclusion, such as $^{13}$C, are symmetrical but do not contain a chiral plane of symmetry in which protons reflect neutrons on opposite sides of the plane.

# 4. Discussion

Since the size of the nucleus is arguably the most fundamental intrinsic physical property, correlation between the predicted radius and the experimental charge radius is a primary figure of merit. Alternating nucleon model radius predictions demonstrate strong positive correlation. An ad hoc association or arrangement of nucleons, with no regard to a set of guiding principles, might randomly generate radius predictions that correlate with experimental. But the nucleon structures featured in figure 6 were assembled using first principles including empirically accepted nucleon size, shape, and separation distance, guided by the known stability of the proton-neutron *NN* interaction and the observed stability of a ratio of neutrons to protons greater equal to or greater than one. The final step in structural elucidation was vetting a structure's calculated predicted radius against the experimental charge radius. The parameters of inclusion were sufficiently constraining that generally only one structure per nuclide met or optimized criteria. The result was a mix of symmetrical and asymmetrical structures, and bilateral structural symmetry emerged as a sensitive and specific predictor of light nuclide abundance.

There is a distinction between the concept of the resilience of a nuclide structure, or its resistance to destruction, as compared to thermodynamic stability, which relates to the internal energy of the products of a reaction relative to the reactants ($\Delta$ H). The nucleon binding energy is an indicator of *thermodynamic* stability, which generally increases from the lightest nuclides through $^{56}$Fe (or more exactly $^{62}$Ni, which is the most tightly bound nuclide with a binding energy of 8.8 MeV/nucleon [16]). Whereas *resilience* of an individual nucleus relates instead to its ability to resist destruction in the face of a dynamic or kinetic impact load. For example, the spallation nuclides are formed and then destroyed during stellar nucleosynthesis [19]. They have a low relative threshold of destruction within the hyperkinetic environment of a supernova, for instance. Conversely, the hypothesis here is that the equinucleonic nuclides resist destruction in the highly kinetic environment of stellar nucleosynthesis, and thus emerge in greater relative abundance. It may be useful to think of the 'threshold of destruction' in this context as equivalent thermodynamically to the reaction activation energy $E_a$, which is independent of $\Delta$ H. Thus, while $^{56}$Fe is highly thermodynamically stable, the equinucleonic nuclides are 'stable' in the sense that their abundance suggests that they are resistant to destruction by kinetic impact.

### 4.1. Steric selection anomalies

There are seven (out of 33) instances where the model predicts the formation of the next stable nuclide through the addition of a proton, but instead, the addition of a neutron produces the next stable nuclide. For example, the deuteron adds a proton to produce stable $^{3}$He, whereas the model anticipates the addition of a neutron to form a triton (which is unstable). In another example, the model's steric selection mechanism anticipates the addition of a proton to $^{7}$Li to produce $^{8}$Be, but $^{8}$Be is unstable. Likewise, the model anticipates the addition of a proton to $^{17}$O, $^{21}$Ne, $^{25}$Mg, $^{29}$Si, and $^{33}$S to form (respectively) $^{18}$F, $^{22}$Na, $^{26}$Al, $^{30}$P, and $^{34}$Cl (all of which are unstable). Instead, the stable product arises from the addition of a neutron resulting in the stable isobars $^{18}$O, $^{22}$Ne, $^{26}$Mg, $^{30}$Si, and $^{34}$S. It may be the case that the predicted nuclide formed but decayed shortly after formation, while the stable isobar represents the scarce accumulation of a product of a minor side reaction. Nonetheless, the model's steric selection mechanism predicts the next stable nuclide in 26 of the 33 instances. Importantly, the nuclides resulting from these 26 successful predictions constitute 99.39% of the ordinary polynucleonic matter of the universe by mass, while the seven nuclides resulting from the unanticipated addition of a neutron constitute a scant 0.09% of the polynucleonic mass of the universe.

### 4.2. A potential energy barrier arises between regularly alternating and unequal charge sequences

1.1.1 The Coulomb barrier occurs at the quantum interface between the strong and electromagnetic forces as a pair of nuclei approach each other. The electromagnetic force predominates beyond the barrier owing to the repulsion between nuclei. Within the Coulomb barrier, however, intense attraction confines nucleons and quarks to the atomic nucleus.

$$U = k\frac{q_1 q_2}{r} \tag{2}$$

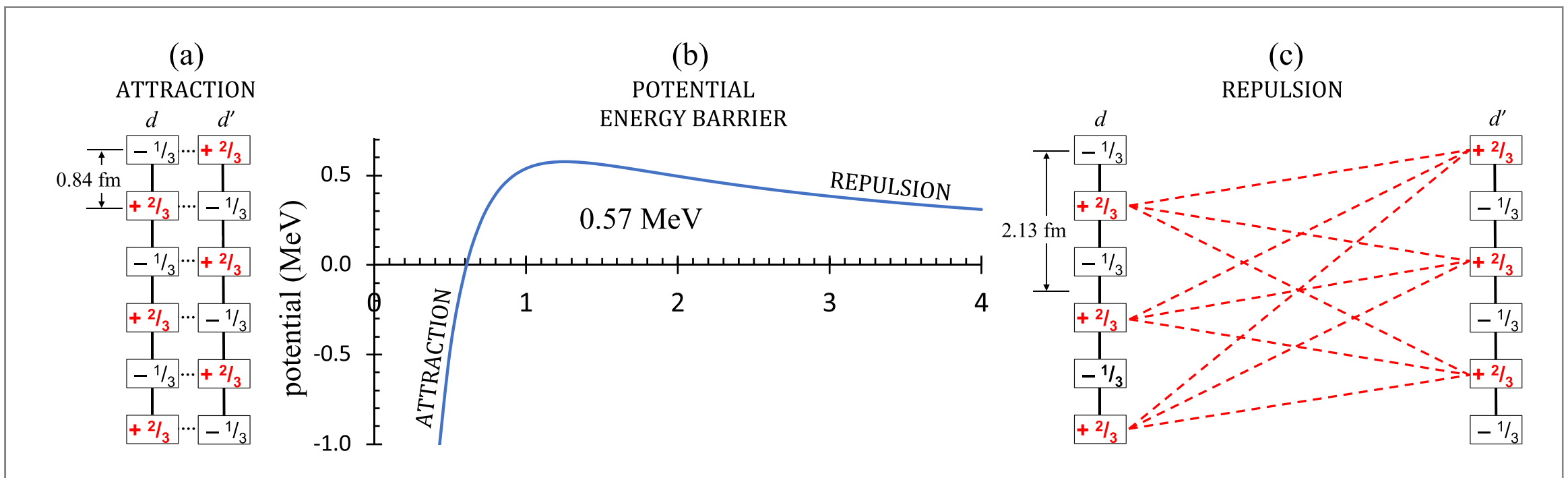

**Figure 11.** Consequences of alternating/unequal charge geometry. (a) A pair of opposing charge sequences, each with six alternating/unequal charges, demonstrate (a) short-range attraction, (c) far-range repulsion, and (b) a potential barrier between.

The atomic nucleus contains protons and neutrons, each with three fractionally charged quarks: The proton with two $+2/3$e up quarks and a $-1/3$e down quark, the neutron with two down quarks and an up quark. The proton-neutron bound state (deuteron) is a linear sequence of six regularly alternating average quark positions as shown in figure 4(c). The fusion potential between a pair of deuterons can be modelled by calculating the individual Coulomb potentials (Formula 1) between the six charges on the approaching deuteron versus six charges on the target deuteron at a series of incremental distances up to 4.0 fm, as shown in figure 11. The potentials are summed within matrices, and the potential/distance data plotted to produce a fusion potential curve. The positive charge on each nucleus results in repulsion at far-range. However, because the charges regularly alternate at a distance of 0.84 fm apart, the $+2/3$e charges on one deuteron sequence can align with $-1/3$e charges on the other at close-range, and the proximity of opposite charges results in strong attraction. The model predicts a Coulomb barrier height of 0.57 Mev at 1.3 fm.

Proton-deuteron (*p-d*) fusion generates a similar fusion potential curve profile using a linear regularly alternating sequence of the proton's three quark charges, but predicts a proton-deuteron Coulomb barrier height of 0.88 Mev at 0.9 fm. Within the model, the neutron's three quark charges demonstrate strong near-range attraction during neutron-deuteron (*n-d*) fusion (arising from opposite quark alignment) but negligible far-range repulsion (Coulomb barrier height of 0.00946 MeV at 1.8 fm) owing to the net zero charge on the neutron, consistent with empirical knowledge of neutron behavior.

For the purpose of this example, the alternating and unequal charges selected are equivalent to the charges on the up-and-down-quark. But the phenomenon is generalizable to any equally-spaced opposite/unequal charge sequence or matrix [40]. In the example of quarks, the positive charge magnitude is exactly double the negative charge. But a potential barrier would also arise if the positive charge magnitude were three- or four-times negative charge within a sequence or matrix of alternate charge, or, for that matter if the negative charge were greater than the positive charge. In fact, the method extends to alternating and unequal magnetic fields as well, which is no surprise given the unification of electricity and magnetism within the electromagnetic fundamental force [41]. A pair of magnetic arrays, each comprising a sequence of equally-spaced N/S magnets, wherein N magnets exhibit twice the pull-force of S magnets, demonstrate a fusion-like force/distance curve and a magnetic potential barrier analogous to the Coulomb barrier [42].

## 5. Conclusion

There is keen interest in the mechanics and parameters of nuclear fusion, including methods of modelling the Coulomb barrier for scientific and educational purposes. The Coulomb barrier is a type of potential energy barrier that results from the interplay of two fundamental interactions: the strong interaction at close-range within $\approx$1 fm, and the electromagnetic interaction at far-range beyond the Coulomb barrier. The microscopic range of the strong interaction, on the order of one femtometre, makes it challenging to model and few classical examples exist on the human scale [43]. Here we have shown that regularly-spaced opposite and unequal electrostatic point charges (equivalent to the charges on up and down quarks) possess the capacity to model an electrostatic potential energy barrier.

The alternating nucleon model predicts a variety of other nuclear phenomena. The nucleon configuration of a given nuclide sterically determines the sequence, isospin (proton vs. neutron), and occupancy of the next added nucleon. The presumption of regular quark separation enables the use of the radius of the regular polygon formula to accurately predict nuclide radii. Resulting light nuclide structures exhibit loose, group-specific periodicity in the nucleon separation energy and nuclear magnetic moments that corresponds to the physical structure of the nucleon configuration within a group's anisotrope. Perhaps more trivially, the model predicts the instability of a 5 amu nuclide, and the proposed ring structures of the helium isotopes, $^{12}$C, and $^{16}$O are consistent with the experimental finding of a central depression in the point-proton charge densities.

The vast majority of ordinary baryonic matter (99.5%) comprises just eight nuclides: $^{4}_{2}$He, $^{12}_{6}$C, $^{14}_{7}$N, $^{16}_{8}$O, $^{20}_{10}$Ne, $^{24}_{12}$Mg, $^{28}_{14}$Si, and $^{32}_{16}$S (aside from hydrogen). Aside from their superabundance, the second most striking feature of the Big Eight is that they all contain equal numbers of protons and neutrons: they are equinucleonic ($Z = N$). Given their overwhelming abundance, the relevance of the Big Eight to a complete understanding of nuclear structure cannot be overstated.

Symmetry is ubiquitous in nature, where evidence abounds that structural symmetry confers structural stability (resistance to destruction) [44–47]. Within the alternating nucleon model, bilateral structural symmetry develops within cylindrical lattice structures as a sensitive and specific predictor of nuclide natural abundance, and structural *asymmetry* emerges as a predictor of cosmic scarcity. We conclude that matter aggregates in alternating nucleon sequences to optimize the formation of stable proton-neutron interactions and maximize the number of *pn* SRC pairs. Nuclides having equal numbers of protons and neutrons intrinsically contain the maximum number of *pn* SRC pairs, and the alternation of nucleons implicitly generates the observed equal numbers of protons and neutrons within the most abundant forms of matter. Within the model, the equality of proton/neutron numbers manifests as structural symmetry, conferring to the eight superabundant equinucleonic nuclides sufficient structural resilience to withstand highly exothermic Big Bang or stellar nuclear reaction conditions and emerge in high relative cosmic abundance.

## Acknowledgments

A hardy thanks to Tim Lohof for his unwavering belief that nuclear quarks alternate, for his countless hours of proofreading while pursuing a doctorate in history at the Geneva Graduate Institute. Thanks to Dr Brian Hadley Reed, USAF Ret., for his steady moral and technical support. Thanks to Pat Callis, my quantum chemistry and professor emeritus at Montana State University, for giving me his time and patience. Also, thanks to Daniel J Glenn, principal architect at 7 Directions Architecture for his insights regarding structural symmetry and its role in structural integrity. Special thanks to my late son Christopher Khalil for his wisdom, patience, and insight.

## Data availability statement

All data that support the findings of this study are included within the article (and any supplementary files).

## ORCID iDs

Ray Walsh https://orcid.org/0000-0003-3887-2966